\documentstyle[graphicx,twocolumn,12pt]{article}

\def\beq{\begin{equation}}
\def\eeq{\end{equation}}
\begin{document}

\pagestyle{empty}
{\onecolumn

\noindent
\small
Reprinted from LALP-92-006, CNLS NEWSLETTER No. 82, September 1992.
Center for Nonlinear Studies, Los Alamos National Lab., Los Alamos NM 87545
\bigskip
\phantom{x}
\vskip 2in
\phantom{x}

\section*{\centerline{Patterns that Grow (in a dish)}}

\medskip
{\it \centerline{William J. Bruno}}
\bigskip
\subsection*{\centerline{Abstract}}

Chemotactic bacteria have been observed to congregate
into highly regular patterns.  
When the bacteria are placed in the center of a dish,
a wave of bacteria can travel outward, leaving a
regular pattern of spots or stripes in its wake.
Although chemotaxis and excretion of an attractant can
readily cause a pattern forming instability from
a uniform state, they are not capable of generating
patterns starting from a single spot.  These patterns are
apparently formed with the help of bacterial growth
and depletion of nutrients in the growth medium.

\vfill
}

\twocolumn
\setcounter{page}{2}

Recent experiments  have found that
under certain special conditions, bacteria will form
very regular patterns in a Petri dish [Budrene \&~Berg, 1991].  These 
patterns include equally spaced radial stripes, radial columns of
spots, sunflower-like arrays of spots, and spots with radial tails
arranged in chevrons.  Bacteria were initially added to the center
of the dish, which contained a growth medium suspended in semi-solid
agar.  The patterns formed in the wake of a
circular ``wave'' radiating from the center of the dish.
Less regular patterns could be obtained starting from bacteria
spread uniformly in a thin layer of liquid.  It was shown that the bacteria
are capable of excreting aspartate, to which bacteria are 
strongly attracted.

It is obvious that if cells swim towards an attractant that they
themselves excrete, the cells will tend to form clusters, since
once cells become over abundant in one region, nearby cells will
be attracted to that same region.  In fact, this may be the most
intuitive example of a uniform state that is unstable to
perturbations that break the spatial symmetry.   Yet we shall see
that getting patterns of the type observed is not as easy 
as one might think.  This will lead to ideas about why the bacteria
are behaving in such a peculiar way.

\subsection*{\normalsize A simple model}
As is typical
in pattern formation problems, determining exactly
what pattern will form depends on nonlinearities in the problem,
but a lot can be learned from a linear analysis as well.
So, let us begin with the linear stability analysis of the
simplest possible model for such an instability:
\begin{eqnarray}
\dot{B} &=& D_B \nabla^2 B - R \nabla \cdot (B \nabla \!A)\nonumber\\
\dot{A} &=& D_{\!A} \nabla^2 \!A + K B - \mu A
\end{eqnarray}
Here $B$ is the concentration of bacteria, $A$ the concentration of
attractant, $D_B$ and $D_{\!A}$ are diffusion coefficients, $R$ measures
the attractiveness of the attractant, $K$ is the rate at which bacteria
excrete the attractant, and $\mu$ determines the lifetime of $A$.
Equations of this type have been considered in the
context of embryonic bone formation [Oster \& Murray, 1989].
In what follows, $\mu$ will be assumed to be negligibly small.

The only nonlinearity in equation (1) is the quadratic $B \nabla \!A$ term.
By linearizing about a uniform steady state with a concentration
$B_0$ of bacteria, one finds that the system is neutrally stable
to uniform perturbations and unstable to long wavelength
perturbations, i.e., those with wavenumber $k$ such that 
$0<k<\sqrt{R K B_0/ D_B D_{\!A}}$.  The maximally unstable mode
is near the middle of the unstable band and grows with a rate
proportional to $(D_B+D_{\!A}) R K B_0/D_B D_{\!A}$.  That $R$ and $K$
appear together in the numerator of these expressions was to 
be expected, since they determine the rates of the two processes
which are intuitively responsible for the instability.
We can conclude that our intuition was correct, and that
as long as the physical dimensions of the system (e.g., the size
of the Petri dish) is larger that $\sqrt{D_B D_{\!A}/ R K B_0} $ then
bacteria in an initially uniform state will
form a pattern, consistent with the experiments done in the liquid
medium.

The fact that $R$ and $K$ appear in the above expressions only
as the product $R K$ has an interesting interpretation which
may be of some biological relevance.  Because only the product
appears, the linear analysis will give exactly the same results
if each constant is multiplied by $-1$.  If $R$ is negative, then
$A$ is a repellent, not an attractant.  If $K$ is negative, then
$A$ is destroyed by the bacteria, not created.  Thus, at least
at the linearized level, being repelled by something you destroy
is equivalent to be attracted to something you make.  Either
situation results in congregation.

Indeed, being repelled by something the bacteria destroy
has been proposed as the underlying reason for these
patterns [Budrene \& Berg].  Since the carbon source used
in the experiments is one of the more oxidized intermediates
of the Krebs cycle, it is possible that the bacteria are
in a state of oxidative stress.  Since the bacteria use up
oxygen, it makes sense that by clustering together the bacteria
could lower the local oxygen concentration and thereby relieve
their stress.  The intuition for pattern formation is the
same as before, but vice versa:  fluctuations that lower
the concentration of bacteria in a region result in fewer bacteria
in that region and hence more repellent, and so forth.
If the oxidative stress idea is correct, it means that bacteria
are making use of this symmetry.  Rather than swimming  away from
the oxygen they consume, they simply generate an attractant
and swim towards it.  The effect is the same.

\subsection*{\normalsize Patterns from a point}
Understanding the patterns that form from a single initial spot
is a bit more difficult; for one thing, we cannot perturb about a 
uniform steady state.  Let us begin by carrying out 
a dimensional-type argument.
For a spot to be at a steady state, the flux outward due to diffusion
must be offset by the flux inward due to the attractant:
\beq
D_B \nabla B  = R B \nabla \!A.
\eeq
  At steady state, the
outward flux of $A$ must be balanced by excretion:
\beq
D_{\!A} \nabla \!A = D_B B_T /r^{d-1},
\eeq
(ignoring  multiples of $\pi$)  where $r$ represents the radius of the
spot, $B_T$ is the total amount of
bacteria in the spot, and $d$ is the number of spatial dimensions
in the problem.  Putting these two equations together and estimating
$\nabla B$ as $B/r$  we find
that
\beq
r^{d-2} = {R K B_T \over D_B D_{\!A}}.
\eeq
In one dimension ($d=1$) this says $r= D_B D_{\!A}/RKB_T$, and
we can put this equation to the test
because the equation  for a steady state is exactly soluble.
Substituting $B=-D_{\!A}\nabla^2 \!A/k$ yields:
\beq
-D_B\nabla^4 \!A - R \nabla(\nabla^2 \!A \nabla \!A) = 0
\eeq
which can be repeatedly integrated, and the solution
for $B$ is
\beq
B(x)= {R K B_T^2 \over 8 D_B D_{\!A}} \hbox{{\rm sech}}^2({R K B_T x \over 4 D_B D_{\!A}}).
\eeq
We see that our estimate for the size of the spot was quite good.
Also, we have learned that in one dimension, a steady state can
have at most one spot.  However, since the spots decay exponentially
we can assume that a state with several spots very far apart will
take a very long time to relax to a single spot.

\subsection*{\normalsize Peculiarities of 2D}
In two dimensions, the situation is different.  First of all, the
equation for a steady state does not readily yield exact solutions.
Secondly, our equation for the radius of the spot is obviously 
meaningless.  If we carry out the same argument again, taking into
account that all the $r$s cancel, we realize that a steady state
should occur only when $B_T$, the total amount of bacteria, equals
a certain value, proportional to
\beq
B_T= {D_B D_{\!A} \over K R}.
\eeq
If there are more than this many bacteria in a spot, it should collapse
to a singularity.  If there are fewer, the spot will spread.

If the spot spreads, we might hope that it will spread into a fairly
uniform state, which would then be ripe for undergoing pattern
formation as discussed above.  How far does it have to spread
for this to happen?  The size of the spot must be bigger than
the critical wavelength of $\sqrt{D_B D_{\!A} /R K B_0}$, and in fact
should be much bigger if we are to get
any kind of non-trivial pattern.  Naturally $B_0$ will go as $B_T/r^2$,
so we find
\beq
r \gg \sqrt{D_B D_{\!A} \over R K B_T}\,\, r
\eeq
or,
\beq
{R K B_T \over D_B D_{\!A}} \gg 1.
\eeq
Thus, there is {\em no} $r$ at which patterns begin to form!  If the
quantity $RKB_T/D_B D_{\!A}$ is bigger than one, we already found that
the single initial spot will collapse, not spread.  If it is less than
one, the initial spot will spread and keep on spreading.  
This has been confirmed by numerical
simulations, starting with a Gaussian spot.  Depending on the initial
shape, it is possible to get pattern formation within the original
spot, but it is not possible for the spot to spread and leave a pattern
in its wake, as seen in the experiments.

\subsection*{\normalsize What makes the wave?}
We see that the formation of patterns of isolated spots 
starting from a single central spot requires more than 
just changing the initial conditions of a system whose uniform
state is unstable.  Thus it appears likely that the experiments
with  bacteria involve more than just attraction towards an
excreted molecule.  

More evidence that this process cannot be described by the
equations (1) alone comes from the observation that the circular
wave that travels outward from the center of the dish moves
at a constant velocity.  Such wave-like solutions in diffusive media
are typically fronts at the boundary between two quasi-steady states.
In this case the  steady state ahead of the wave front is obviously the
absence of any bacteria.  The steady state behind the front
probably consists of bacteria whose growth has stalled due to 
depletion of some nutrient in the growth medium.

This is consistent with the fact that the patterns seem to have
the same fundamental wavelength everywhere; the spacing of spots
or stripes is the same  at the outer edge of the dish as
it is near the center.  From our experience with equation (1), this
hints that the concentration of bacteria in the wake of the wave
is unchanged as the wave travels outward, since the critical wavelength
depends on the concentration of bacteria.  The concentration of
bacteria in the wake would indeed be constant if the wave represents
a transition from no bacteria to bacteria whose growth using the
more readily metabolized nutrients has saturated.

This idea makes sense in terms of the biochemistry of the
Krebs cycle as well.  In the experiments, the bacteria
excrete large amounts of aspartate; an intracellular concentration
of  0.2M  would be necessary if the bacteria are simply excreting
aspartate they have stored.  Thus, it is likely they are making
it from the Krebs cycle intermediates that are abundant in the
growth medium that was used.  When the leucine and threonine in
the growth medium are used up, the bacteria will be unable to catabolize
the remaining amino acids according to the normal Krebs cycle.
Under these conditions, one would expect oxaloacetate to accumulate
in the cell, and this would tend to cause creation of aspartate through the
action of aspartate-glutamate transaminase [Lehninger, 1970].  On the
other hand, when
leucine or threonine is abundant, one expects bacteria to consume 
any available aspartate.

Presumably, the bacteria grow much faster when leucine or threonine
is available than when not, and while growing they will consume these
amino acids rapidly.  Considering these two processes plus diffusion
leads us to consider the equations:
\begin{eqnarray}
\dot{B} &=& D_B \nabla^2 B + K_{\hbox{\scriptsize grow}} B L \nonumber \\
\dot{L} &=& D_L \nabla^2 L - K_{\hbox{\scriptsize eat}} B L
\end{eqnarray}
where $L$ is the concentration of leucine or threonine (or their sum).
This set of equations does admit wave-like solutions such as those
seen in the experiments.  Ahead of the wave, $B$ is zero; behind it,
$L$ is (nearly) zero.  In fact, in the special case where 
$K_{\hbox{\scriptsize grow}} D_L = K_{\hbox{\scriptsize eat}} D_B$
then $L=L_0 - K_{\hbox{\scriptsize eat}} B/K_{\hbox{\scriptsize grow}}$
and one has
\beq
\dot{B} = D_B \nabla^2 B + K_{\hbox{\scriptsize grow}} B L_0 - K_{\hbox{\scriptsize eat}} B^2,
\eeq
with $L_0$ the initial concentration of leucine.
This is known as Fisher's equation, and has propagating wave solutions
with a constant speed equal to $\sqrt{4D_B K_{\hbox{\scriptsize grow}} L_0}$
[Fisher, 1937].

The value of $D_B$ was estimated by the experimenters to be 
$4.8\times 10^{-6}\rm~cm^2/sec$,
and since the medium is designed to saturate the rate of growth, 
$K_{\hbox{\scriptsize grow}} L_0$ should be in the range of the maximal
growth rate for bacteria.  Taking a doubling time of 30 minutes, we get
an estimated wave speed of $8.6\times 10^{-5}~\rm cm/sec$.  This estimate
is only a factor of four larger than
the wave speed of $2.1\times 10^{-5}~\rm cm/sec$ quoted in the experimental
paper.   Thus it is plausible that bacterial growth and nutrient depletion 
are the main factors responsible for the observed wave.

The experiments done with  a pre-stirred bacteria/growth medium mixture
showed that patterns could form in less than a generation time, and
that therefore growth was not necessary for pattern formation.  This
is consistent with the ideas presented here.  Getting patterns to
form from a uniform state is ``easier'' than getting them from a single
initial spot.  It is likely that the traveling wave, which depends
on growth, sets up conditions in its wake that are similar to those
in the stirred medium.  Once these conditions are established, the
attractive instability takes over and patterns form.  

The figures show some results of numerical simulations of  model equations.
Many models were tried; the most convincing patterns formed in those
that had bacteria, aspartate, and ``leucine'' (which could really be any
nutrient in short supply) as the main variables.  In all of the models
that produced a wave, bacteria grew and ate leucine as in the above
equation, and produced aspartate in the presence of succinate, which
maintained a nearly constant concentration.  Models where the bacteria
were chemotactically attracted to both aspartate and leucine gave waves
where the bacteria concentration was higher near the wave front
than in the wake, as was also seen in the experiments.  

In no model has the striking geometric regularity of the bacterial
patterns been seen.  This could be due to a tendency to use parameter
values that will allow the wave to be followed an appreciable distance
without using vast amounts of computer time.  Forcing the patterns to
form quickly probably makes them more random.  Of course, pattern selection
depends very much on nonlinearities.  Some of the nonlinearities in
this problem---such as receptor saturation---can be (and were) readily modeled
at a  qualitative level, but there are likely to be nonlinear effects in
the chemistry (think of many enzymes, especially the allosteric ones, 
in the Krebs cycle) and elsewhere in the system that even the most
ambitious model would fail to capture.  Still, the simple analysis that
was presented here leads to some suggestive possibilities in terms
of why the bacteria are doing what they do.  The question of whether
the bacteria are being smart and avoiding stress or just making
patterns for no useful reason at all remains unanswered.

\subsection*{\normalsize Acknowledgments}
Micah Dembo has provided valuable advice on the numerical and biological
aspects of this problem;  Mark Mineev helped in gaining insight into
the equations; and John Pearson also contributed greatly to my understanding
of this problem.  I also wish to thank the Santa Fe Institute for its
hospitality, and the Advanced Computing Laboratory for use of the TMC CM-200, and Susan Coghlan and Stephen Pope for their technical support.

\subsection*{\normalsize References}
\noindent  Budrene, E. O. \& H. C. Berg, {\it Nature} {\bf 349}, 630--633 (1991).

\noindent Fisher, R. A., {\it Eugenics} {\bf 7}, 355--369 (1937).

\noindent Oster G. F. \& J. D. Murray, {\it J. Exp. Zool.} {\bf 251}, 186--202 (1989).

\noindent Lehninger, A. L., {\it Biochemistry} (Worth Publishers, New York, 1970).

\onecolumn
\includegraphics{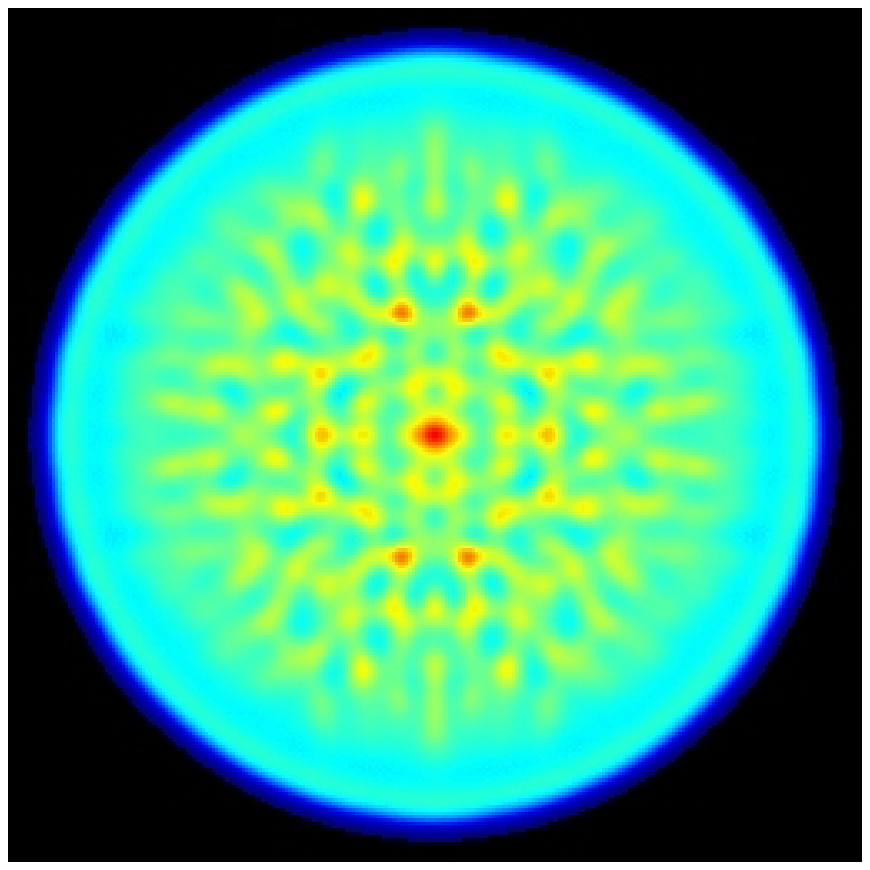}

\medskip

\noindent Figure 1.: A simulation of a pattern forming from a 
small spot.  The lighter, circular region denotes higher bacteria
concentration.  The region gets larger as time progresses.  This
image was converted from color; the small dark spots in the bright region
actually contain more bacteria than the surroundings, while the black
outer region contains almost no bacteria.  This figure and the others
were made by pasting together four copies of a simulation of a quarter
circle wedge.

\bigskip
\vfill \eject
{\samepage
\includegraphics{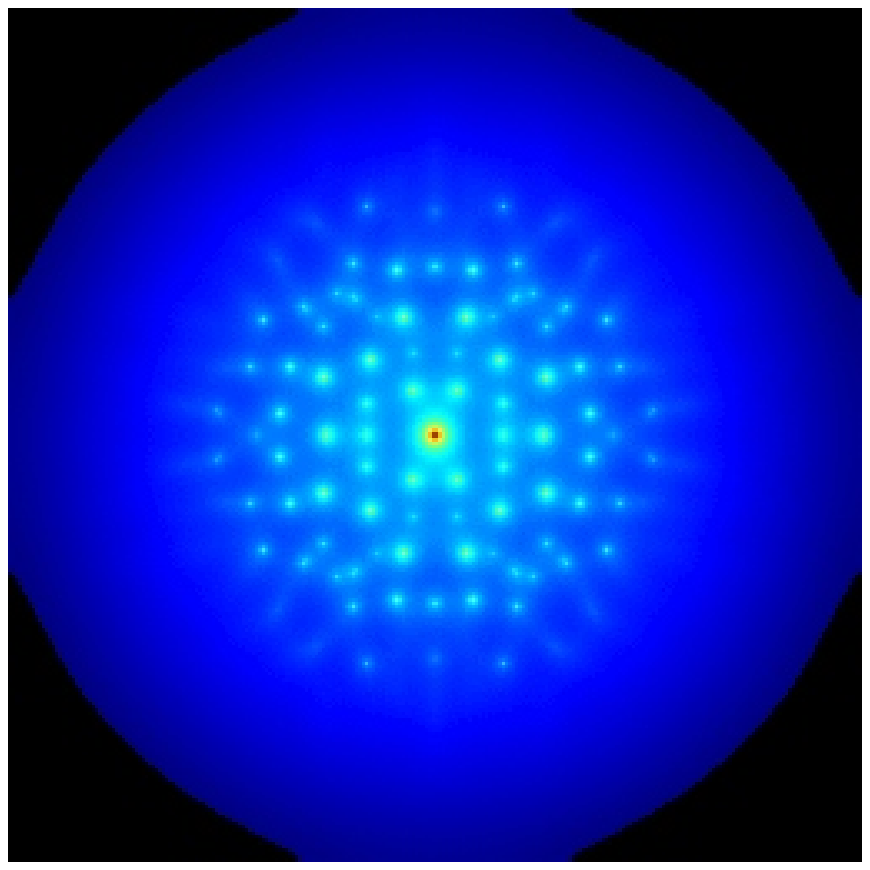}

\medskip

\noindent Figure 2.: With these parameters, the spots are more localized,
as in the real experiments, but the arrangement of the spots seems
fairly irregular compared with what the real bacteria did.
}

\bigskip
\vfill \eject
{\samepage
\includegraphics{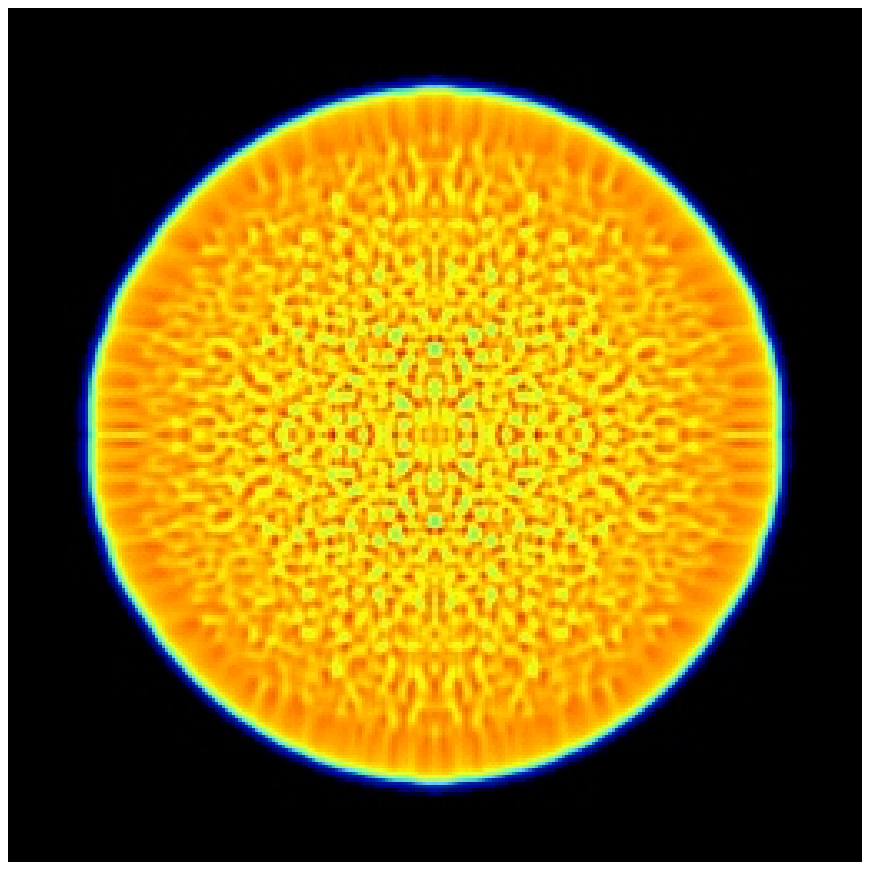}

\medskip

\noindent Figure 3.:  In this simulation, the bacteria were chemotactically
attracted to the leucine as well as the aspartate.  The predominant wavelength 
shorter because a different scale was used.  
}

\bigskip
\vfil

\end{document}